\documentclass[conference, 10pt]{IEEEtran}
\usepackage[T1]{fontenc}
\usepackage[utf8]{inputenc}
\usepackage{amssymb}
\usepackage[pdftex]{graphicx}
\usepackage{dsfont}
\usepackage{siunitx}
\usepackage{mathtools}
\usepackage{mathabx}
\usepackage{array}
\usepackage[noadjust]{cite}
\usepackage{glossaries}
\usepackage{algorithm}
\usepackage{algpseudocode}
\usepackage{xcolor}
\usepackage{tikz}
\usepackage{pgfplots}
\usepackage{balance}
\usepgfplotslibrary{fillbetween} 
\usepackage[noadjust]{cite}
\usepackage{hyperref}

\usetikzlibrary{decorations.pathreplacing,calc}

\makeatletter
\let\MYcaption\@makecaption
\makeatother

\usepackage[font=footnotesize]{subcaption}

\makeatletter
\let\@makecaption\MYcaption
\makeatother
\algnewcommand{\LineComment}[1]{\(\triangleright\) #1}

\newacronym{ACM}{ACM}{adaptive coding and modulation}
\newacronym{ADC}{ADC}{analog-to-digital conversion}
\newacronym{AGC}{AGC}{automatic gain control}
\newacronym{AWGN}{AWGN}{additive white Gaussian noise}
\newacronym{BER}{BER}{bit error rate}
\newacronym{BS}{BS}{base station}
\newacronym{BLER}{BLER}{block error rate}
\newacronym{BP}{BP}{backpropagation}
\newacronym{BPTT}{BPTT}{backpropagation through time}
\newacronym{CE}{CE}{cross-entropy}
\newacronym{CFO}{CFO}{carrier frequency offset}
\newacronym{CSI}{CSI}{channel state information}
\newacronym{DAC}{DAC}{digital-to-analog conversion}
\newacronym{DL}{DL}{deep learning}
\newacronym{DFT}{DFT}{discrete Fourier transform}
\newacronym{FFT}{FFT}{fast Fourier transform}
\newacronym{GAN}{GAN}{generative adversarial network}
\newacronym{GRU}{GRU}{gated recurrent unit}
\newacronym{iid}{i.i.d.\@}{independent and identically distributed}
\newacronym{IFFT}{IFFT}{inverse fast Fourier transform}
\newacronym{KL}{KL}{Kullback-Leibler}
\newacronym{LLR}{LLR}{log likelihood ratio}
\newacronym{LSTM}{LSTM}{long short-term memory}
\newacronym{LMMSE}{LMMSE}{linear minimum mean squared error}
\newacronym{MDP}{MDP}{Markov decision process}
\newacronym{ML}{ML}{machine learning}
\newacronym{MLP}{MLP}{multilayer perceptron}
\newacronym{MIMO}{MIMO}{multiple-input multiple-output}
\newacronym{MSE}{MSE}{mean squared error}
\newacronym{NN}{NN}{neural network}
\newacronym{OFDM}{OFDM}{orthogonal frequency-division multiplexing}
\newacronym{pdf}{pdf}{probability density function}
\newacronym{pmf}{pmf}{probability mass function}
\newacronym{QPSK}{QPSK}{quadrature phase-shift keying}
\newacronym{PSNR}{PSNR}{Peak Signal to Noise Ratio}
\newacronym{RBF}{RBF}{Rayleigh block-fading}
\newacronym{ReLU}{ReLU}{rectified linear unit}
\newacronym{RTN}{RTN}{radio transformer network}
\newacronym{RL}{RL}{reinforcement learning}
\newacronym{RNN}{RNN}{recurrent neural network}
\newacronym{SFO}{SFO}{sampling frequency offset}
\newacronym{SER}{SER}{symbol error rate}
\newacronym{SNR}{SNR}{signal-to-noise ratio}
\newacronym{SINR}{SINR}{signal-to-interference-plus-noise ratio}
\newacronym{SGD}{SGD}{stochastic gradient descent}
\newacronym{wrt}{w.r.t.\@}{with respect to}
\newacronym{ELU}{ELU}{exponential linear unit}

\renewcommand{\vec}[1]{\mathbf{#1}}

\newcommand{\bv}{\vec{b}}

\newcommand{\ev}{\vec{e}}

\newcommand{\hv}{\vec{h}}

\newcommand{\nv}{\vec{n}}

\newcommand{\sv}{\vec{s}}

\newcommand{\xv}{\vec{x}}
\newcommand{\yv}{\vec{y}}
\newcommand{\zv}{\vec{z}}
\newcommand{\zerov}{\vec{0}}

\newcommand{\Am}{\vec{A}}

\newcommand{\Cm}{\vec{C}}
\newcommand{\Dm}{\vec{D}}

\newcommand{\Hm}{\vec{H}}
\newcommand{\Id}{\vec{I}}

\newcommand{\Qm}{\vec{Q}}
\newcommand{\Rm}{\vec{R}}

\newcommand{\Um}{\vec{U}}

\newcommand{\Xm}{\vec{X}}

\newcommand{\Cc}{{\cal C}}

\newcommand{\Lc}{{\cal L}}

\newcommand{\Nc}{{\cal N}}

\newcommand{\Xc}{{\cal X}}

\newcommand{\CC}{\mathbb{C}}

\newcommand{\RR}{\mathbb{R}}

\newcommand{\LB}{\left(}
\newcommand{\RB}{\right)}

\renewcommand{\exp}[1]{\mathop{\mathrm{exp}}\LB #1\RB}

\newcommand{\EE}{{\mathbb{E}}}

\begin{document}
\title{Deep HyperNetwork-Based MIMO Detection}
\author{
\IEEEauthorblockN{Mathieu Goutay\IEEEauthorrefmark{1}\IEEEauthorrefmark{2}, Fayçal Ait Aoudia\IEEEauthorrefmark{1}, and Jakob Hoydis\IEEEauthorrefmark{1}}
\IEEEauthorblockA{\IEEEauthorrefmark{1}Nokia Bell Labs, Paris-Saclay, 91620 Nozay, France}
\IEEEauthorblockA{\IEEEauthorrefmark{2}Univ Lyon, INSA Lyon, Inria, CITI,  69100 Villeurbanne, France \\
mathieu.goutay@nokia.com, \{faycal.ait\_aoudia, jakob.hoydis\}@nokia-bell-labs.com
}}
\maketitle

\begin{abstract}

Optimal symbol detection for \gls{MIMO} systems is known to be an NP-hard problem. 
Conventional heuristic algorithms are either too complex to be practical or suffer from poor performance.
Recently, several approaches tried to address those challenges by implementing the detector as a deep neural network. 
However, they either still achieve unsatisfying performance on practical spatially correlated channels, or are computationally demanding since they require retraining for each channel realization. 
In this work, we address both issues by training an additional \gls{NN}, referred to as the hypernetwork, which takes as input the channel matrix and generates the weights of the neural \gls{NN}-based detector.
Results show that the proposed approach achieves near state-of-the-art performance without the need for re-training.


\end{abstract}

\glsresetall

\section{Introduction} 
\label{sec:introduction}


To keep up with the always increasing mobile user traffic, cellular communication systems have been driven by continuous innovation since the introduction of the first generation in 1979.
The attention is now turning from the fifth to the sixth generation, which some predict should be able to deliver data rates up to 1\:TB/s with high energy efficiency~\cite{roadmap_6G}.
A key enabler is to serve multiple single-antenna users on the same time-frequency resource using a \gls{BS} equipped with a large number of antennas.
However, optimal detection in such \gls{MIMO} systems is known to be NP-hard~\cite{10.1007_s10107-016-1036-0}, and approaches introduced in recent years suffers from unsatisfying performance or become impractical
when the number of antennas or users is large.
Examples of recent approaches include the iterative algorithm AMP~\cite{AMP} or its extension to correlated channels OAMP~\cite{OAMP}.

Recently, advances in \gls{MIMO} detection have been made by using \gls{ML} in conjunction or in place of standard algorithms~\cite{Samsung, DetNet}.
A promising approach is to add trainable parameters to traditional iterative algorithms and interpret the whole structure as a \gls{NN}~\cite{OAMPNet}.
However, these schemes still either suffer from  a performance drop on correlated channels or from high complexity.
One of these approaches is the recently proposed MMNet~\cite{MMNet}, which achieves state-of-the-art performance on correlated channels.
However, it needs to be retrained on each channel realization, which makes its practical implementation challenging.

\begin{figure}
\centering
	\includegraphics[width=0.75\linewidth]{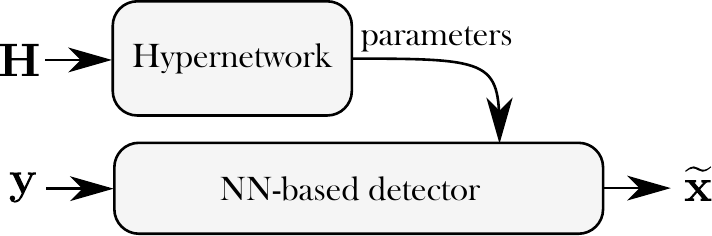}
	\caption{HyperMIMO: A hypernetwork generates the parameters of a \gls{NN}-based detector (MMNet~\cite{MMNet} in this work) }
\label{fig:HG_small}
\end{figure}

In this work, we alleviate this issue by leveraging the emerging idea of \emph{hypernetworks}~\cite{learnet,talking_heads}.
Applied to our setup, it consists in having a secondary \gls{NN}, referred to as the hypernetwork, that generates for a given channel matrix an optimized set of weights for an \gls{NN}-based detector.
This scheme, which we refer to as \emph{HyperMIMO}, is illustrated in Fig.~\ref{fig:HG_small}.
Used with the MMNet detector from~\cite{MMNet}, HyperMIMO replaces the training procedure that would be required for each channel realization by a single inference of the hypernetwork.

We have evaluated the proposed approach using exhaustive simulations on spatially correlated channels.
Our results show that HyperMIMO achieves a performance close to that of MMNet trained for each channel realization, and outperforms the recently proposed OAMPNet~\cite{OAMPNet}.
Our results also reveal that HyperMIMO is robust to user mobility up to a certain point, which is encouraging for practical use.


\textbf{Notations :} 
Matrices and column vectors are denoted by bold upper- and lower-case letters, respectively. 
$x_{i}$ is the $i^{\text{th}}$ element of the vector $\xv$, and $[X]_{i,j}$ the $(i,j)^{\text{th}}$ element of the matrix $\Xm$.
$\text{diag}(\xv)$ is the diagonal matrix composed of the elements of $\xv$, and $\Id_N$ the $N \times N$ identity matrix.
$||\Xm||_F$  is the Frobenius norm of $\Xm$, and $\Xm^H$ its conjugate transpose.

\section{Background} 
\label{sec:Background}

\subsection{Problem formulation}
\label{subsec:problem_definition}
We consider a conventional \gls{MIMO} uplink channel.
We denote by $N_u$ the number of single-antenna users that aim to reliably transmit symbols from a constellation $\Xc$ to a \gls{BS} equipped with $N_r$ antennas.
The channel transfer function is
\begin{equation}
\label{eq:rayleigh}
\yv = \Hm \xv + \nv
\end{equation}
where $\xv \in \Xc^{N_u}$ is the vector of transmitted symbols, $\yv \in \CC^{N_r}$ is the vector of received distorted symbols, $\Hm \in \CC^{N_r \times N_u}$ is the channel matrix, and $\nv \thicksim \Cc\Nc(\mathbf{0}, \sigma^2 \Id_{N_r})$ is the \gls{iid} complex Gaussian noise with power $\sigma^2$ in each complex dimension. 
It is assumed that $\Hm$ and $\sigma$ are perfectly known to the receiver.
The optimal receiver would implement the maximum likelihood detector
\begin{equation}
\label{eq:ml}
\hat{\xv} = \arg \min_{\xv \in \Xc^{N_u}} || \yv - \Hm\xv ||_2^2.
\end{equation}
Unfortunately, solving~(\ref{eq:ml}) is known to be an NP-hard problem due to the finite alphabet constraint $\xv \in \Xc^{N_u}$~\cite{10.1007_s10107-016-1036-0}. 
One well-known scheme is the \gls{LMMSE} estimator which aims to minimize the \gls{MSE} 
\begin{equation}
\label{eq:loss_mmse}
\widetilde{\xv} = \arg \min_{\sv \in \CC^{N_u}} \EE_{\xv, \nv} \left[ || \sv -\xv ||_2^2 \right]
\end{equation}
by restricting to linear estimators.
This allows for a closed-form expression of the solution to~(\ref{eq:loss_mmse})
\begin{equation}
\label{eq:mmse_1}
\tilde{\xv} = (\Hm^H \Hm + \sigma^2 \Id_{N_u})^{-1} \Hm^H \yv.
\end{equation}
Because the transmitted symbols are known to belong to the finite alphabet $\Xc$, the closest symbol is typically selected for each user:
\begin{equation}
\label{eq:mmse_2}
\hat{x}_{i} = \arg \min_{x \in \Xc} || \tilde{x}_{i} - x ||_2^2, \quad \quad \forall i \in \{1, \cdots, N_u\}.
\end{equation}
Although sub-optimal, this approach has the benefit of being computationally tractable. 
Multiple schemes have been proposed to achieve a better performance-complexity trade-off among which \gls{ML}-based algorithms form a particularly promising lead.

\subsection{Machine learning-based MIMO detectors}
\label{subsec:ML-based_algorithms}

\gls{ML} has been leveraged to perform \gls{MIMO} detection in multiple ways.
In~\cite{Samsung}, Chaudhari et al. used an \gls{NN} to select a traditional detection algorithm from a predefined set.
The algorithm with lowest complexity that enables a \gls{BLER} lower than a predefined threshold is chosen.

Another technique is to design an \gls{NN} that performs the detection. 
On example is DetNet~\cite{DetNet} which can be viewed as an unfolded \gls{RNN}.
Although it achieves encouraging results on Rayleigh channels, DetNet's performance on correlated channels is not satisfactory and it suffers from a prohibitive complexity.
In~\cite{DetNet_small}, Mohammad et al. partially addressed this drawback by weights pruning.

A promising approach is to enhance existing schemes by adding trainable parameters.
Traditional iterative algorithms are particularly suitable since they can be viewed as \gls{NN} once unfolded.
Typically, each iteration aims to further reduce the \gls{MSE} and comprises a linear step followed by a non-linear denoising step.
The estimate $\tilde{\xv}^{(t+1)}$ at the $(t+1)$\textsuperscript{th} iteration is 
\begin{equation} 
\label{eq:it_algo}
\begin{aligned}
\zv^{(t)}                 &= \widetilde{\xv}^{(t)} + \Am^{(t)} \left(\yv - \Hm \hat{\xv}^{(t)} + \bv^{(t)} \right)\\
\widetilde{\xv}^{(t+1)} &= \eta^{(t)}\left(\zv^{(t)}, \tau^{(t)} \right)
\end{aligned}
\end{equation}
where the superscript $(t)$ is used to refer to the $t^{\text{th}}$ iteration and $\hat{\xv}^{(0)}$ is set to $\mathbf{0}$.
$\tau^{(t)}$ denotes the estimated variance of the components of the noise vector $\mathbf{z}^{(t)} -\xv^{(t)}$ at the input of the denoiser, which is assumed to be \gls{iid}.
Iterative algorithms differ by their choices of matrices $\Am^{(t)} \in \CC^{N_u \times N_r}$, bias vectors $\bv^{(t)} \in \CC^{N_u}$, and denoising functions $\eta^{(t)}(\cdot)$.
A limitation of most detection schemes is their poor performance on correlated channels. 
OAMP~\cite{OAMP} mitigates this issue by constraining both the linear step and the denoiser. 
OAMPNet~\cite{OAMPNet} improves the performance of OAMP by adding two trainable parameters per iteration, which respectively scales the matrix $\Am^{(t)}$ and the channel noise variance $\sigma^2$.
MMNet~\cite{MMNet} goes one step further by making all matrices $\Am^{(t)}$ trainable and by relaxing the constraint on $\mathbf{z}^{(t)} -\xv^{(t)}$ being identically distributed. 
Although MMNet achieves state-of-the-art performance on spatially-correlated channels, it needs to be re-trained for each channel matrix, which makes it unpractical.

\subsection{Hypernetworks}
\label{HyperNetworks}

Hypernetworks were introduced in~\cite{hypernetworks} as \glspl{NN} that generate the parameters of other \glspl{NN}.
The concept was first used in~\cite{learnet} in the context of image recognition.
The goal was to predict the parameters of a \gls{NN} given a new sample so that it could recognize other objects of the same class without the need for training.
More recently, this same idea was leveraged to generate images of talking heads~\cite{talking_heads}.
In this later work, a single picture of a person is fed to a hypernetwork that computes the weights of a second \gls{NN}.
This second \gls{NN} then generates realistic images of the same person with different facial expressions.
Motivated by these recent achievements, we propose in this work to alleviate the need of MMNet to be retrained for each channel realization using hypernetworks.

\section{HyperMIMO} 
\label{sec:HyperMIMO}

The key idea of this work is to replace the training process required by MMNet for each channel realization by a single inference through a trained hypernetwork.
This section first presents a variation of MMNet which reduces its number of parameters.
The second part of this section introduces the architecture of the hypernetwork, where a relaxed form of weight sharing is used to decrease its output dimension.
Both reducing the number of parameters of MMNet and weight sharing in the hypernetwork are crucial to obtain a system of reasonable complexity.
The combination of the hypernetwork together with MMNet form the HyperMIMO system visible in Fig.~\ref{fig:HG_small}.

\subsection{MMNet with less parameters}
\label{subsec:mmnet}

To reduce the number of parameters of MMNet, we leverage the QR-decomposition of the channel matrix, $\Hm = \Qm \Rm$, where $\Qm$ is an $N_r \times N_r$ orthogonal matrix and $\Rm$ an $N_r \times N_u$ upper triangular matrix.
It is assumed that $N_r > N_u$, and therefore $\Rm = \begin{bmatrix}\mathbf{R_A} \\ \boldsymbol{0}\end{bmatrix}$ where $\mathbf{R_A}$ is of size $N_u \times N_u$, and $\Qm = \left[\mathbf{Q_A} \boldsymbol{Q_B} \right]$ where $\mathbf{Q_A}$ has size $N_r \times N_u$.
We define $\yv^* \coloneq \mathbf{Q_A}^H \yv$ and $\nv^* \coloneq \mathbf{Q_A}^H\nv$, and rewrite~(\ref{eq:rayleigh}) as 
\begin{equation}
\label{eq:QR}
\yv^* = \mathbf{R_A} \xv + \nv^*.
\end{equation}
Note that $\nv^* \thicksim \Cc\Nc(\mathbf{0}, \sigma^2 \Id_{N_u})$.
MMNet sets $\bv^{(t)}$ to $\zerov$ for all $t$ and uses the same denoiser for all iterations, which are defined by
\begin{equation} 
\label{eq:mmnet}
\begin{split}
\zv^{(t)} & = \widetilde{\xv}^{(t)} + \boldsymbol{\Theta}^{(t)} \left( \yv^* - \mathbf{R_A} \hat{\xv}^{(t)} \right) \\
\widetilde{\xv}^{(t+1)} & = \eta \left( \zv^{(t)}, {\boldsymbol{\tau}^{(t)}} \right)
\end{split}
\end{equation}
where $\boldsymbol{\Theta}^{(t)}$ is an $N_u \times N_u$ complex matrix whose components need to be optimized for each channel realization.
The main benefit of leveraging the QR-decomposition is that the dimension of the matrices $\boldsymbol{\Theta}^{(t)}$ to be optimized is $N_u \times N_u$ instead of $N_u \times N_r$, which is the dimension of $\Am^{(t)}$ in~(\ref{eq:it_algo}).
This is significant since the number of active users $N_u$ is typically much smaller than the number of antennas $N_r$ of the \gls{BS}.

The noise at the input of the denoiser $\mathbf{z}^{(t)} -\xv^{(t)}$ is assumed to be independent but not identically distributed in MMNet.
The vector of estimated variances at the $t^{\text{th}}$ iteration is denoted by $\boldsymbol{\tau}^{(t)} \in \RR^{N_u}$ and computed by
\begin{equation} 
\label{eq:noise_std}
\small
\medmuskip=0mu   
\thickmuskip=1mu 
\begin{aligned} 
\boldsymbol{\tau}^{(t)} =\frac{\boldsymbol{\psi}^{(t)}}{N_u} & \left( \frac{||\Id_{N_u}-\boldsymbol{\Theta}^{(t)} \mathbf{R_A}||_{F}^{2}}{||\mathbf{R_A}||_{F}^{2}} \right. \left[||\mathbf{y^*}-\mathbf{R_A} \hat{\xv}^{(t)}||_{2}^{2}-N_{r} \sigma^{2}\right]_{+} \\ &\left.+ ||\boldsymbol{\Theta}^{(t)} ||_{F}^{2} \sigma^{2}\right) \end{aligned}
\end{equation}
where $[x]_{+} = \max(0,x)$, and $\boldsymbol{\psi}^{(t)} \in \RR^{N_u}$ needs to be optimized for each channel realization.
Further details on the origin of this equation can be found in~\cite{OAMP}.
The denoising function in MMNet is the same for all iterations, and is chosen to minimize the \gls{MSE} $\EE_{\xv} \left[||\hat{\xv} - \xv||_2^2 | \zv\right]$ assuming the noise is independent and Gaussian distributed.
This is achieved by applying element-wisely to $(\zv^{(t)}, \boldsymbol{\tau}^{(t)})$
\begin{equation} 
\label{eq:denoiser}
\eta (z, \tau) = \frac{1}{Z} \sum_{x \in \Xc} x \exp{- \frac{|z - x|^2}{\tau} }
\end{equation}
where $Z = \sum_{x \in \Xc} \exp{- \frac{|z - x|^2}{\tau}}$.
MMNet consists of $T$ layers performing~(\ref{eq:mmnet}), and
a hard decision as in~(\ref{eq:mmse_2}) to predict the final estimate $\hat{\xv}$.
One could also use $\widetilde{\xv}^{(T)}$ to predict bit-wise \glspl{LLR}.

\subsection{HyperMIMO architecture}
\label{subsec:implementation_details}

\begin{figure}
\centering
	\includegraphics[width=0.9\linewidth]{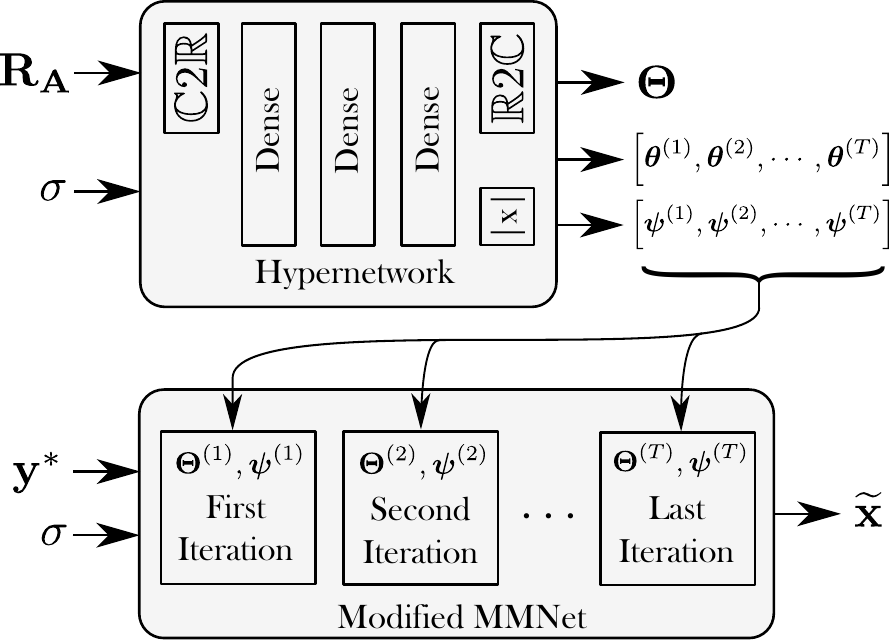}
	\caption{Detailed architecture of HyperMIMO }
\label{fig:HG}
\end{figure} 

Fig.~\ref{fig:HG} shows in details the architecture of HyperMIMO.
As our variant of MMNet operates on $\yv^*$, the hypernetwork is fed with $\mathbf{R_A}$ and the channel noise standard deviation $\sigma$.
Note that because $\mathbf{R_A}$ is upper triangular, only $N_u(N_u+1)/2$ non-zero elements need to be fed to the hypernetwork. 
Moreover, using this matrix as input instead of $\Hm$ has been to found to be critical to achieve high performance.
As detailed previously, the number of parameters that need to be optimized in MMNet was reduced by leveraging the QR-decomposition.
To further decrease the number of outputs of the hypernetwork, we adopt a relaxed form of weight sharing inspired by~\cite{learnet}.
Instead of computing the elements of each $\boldsymbol{\Theta}^{(t)}, t = 1,\dots,T$, the hypernetwork outputs a single matrix $\boldsymbol{\Theta}$ as well as $T$ vectors $\boldsymbol{\theta}^{(t)} \in \RR^{N_u}$.
For each iteration $t$, $\boldsymbol{\Theta}^{(t)}$ is computed by 
\begin{equation}
\label{eq:scaling}
\boldsymbol{\Theta}^{(t)} = \boldsymbol{\Theta} \left( \Id_{N_u} + \text{diag}\left(\boldsymbol{\theta}^{(t)}\right) \right).
\end{equation}
The idea is that all matrices $\boldsymbol{\Theta}^{(t)}$ differ by a per-column scaling different for each iteration.
We have experimentally observed that scaling of the rows leads to worse performance.

Because $\mathbf{R_A}$ is complex-valued, a $\RR2\CC$ layer maps the complex elements of $\mathbf{R_A}$ to real ones, by concatenating the real and imaginary parts of the complex scalar elements.
To generate a complex-valued matrix $\boldsymbol{\Theta}$, a $\CC2\RR$ layer does the reverse operation of $\RR2\CC$.

The hypernetwork also needs to compute the values of the $T$ vectors $\boldsymbol{\psi}^{(t)}$.
Because the elements of these vectors must be positive, a small constant is added and an absolute-value activation function is used in the last layer, as shown in Fig.~\ref{fig:HG}.

HyperMIMO, which comprises the hypernetwork and MMNet, is trained by minimizing the \gls{MSE}
\begin{equation}
\label{eq:loss_hg}
\Lc = \EE_{\xv, \Hm, \nv} \left[ ||\widetilde{\mathbf{x}}_{T}-\mathbf{x}||_{2}^{2} \right].
\end{equation}
Note that this loss differs from the one of~\cite{MMNet}, which is $\frac{1}{T} \sum_{t=1}^T \EE_{\xv, \Hm, \nv} \left[ ||\widetilde{\mathbf{x}}_{t}-\mathbf{x}||_{2}^{2} \right]$.
When training HyperMIMO, the hypernetwork and MMNet form a single \gls{NN}, such that the output of the hypernetwork are the weights of MMNet.
The only trainable parameters are therefore the ones of the hypernetwork.
When performing gradient descent, their gradients are backpropagated through the parameters of MMNet.

\section{Experiments} 
\label{sec:Experiments}

HyperMIMO was evaluated by simulations.
This section starts by introducing the considered spatially correlated channel model.
Next, details on the simulation setting and training process are provided.
Finally, the obtained results are presented and discussed.

\subsection{Channel model}
\label{subsec:channel_model}

\begin{figure}
\center
\includegraphics[width=0.75\linewidth]{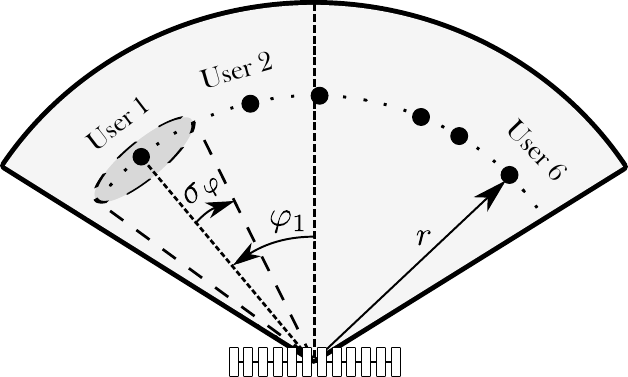}
\caption{Considered channel model. The \gls{BS} has no scatters in its near field, and scattering is only located near users.}
\label{fig:setup}
\end{figure}

The local scattering model with spatial correlation presented in~\cite[Ch.~2.6]{massivemimobook} and illustrated in Fig.~\ref{fig:setup} is considered.
The \gls{BS} is assumed to be equipped with a uniform linear array of $N_r$ antennas, located at the center of a 120$^{\circ}$-cell sector in which $N_u$ single-antenna users are dropped with random nominal angles $\varphi_{u},~u \in \{1, \cdots, N_u \}$.
Perfect power allocation is assumed, leading to all users appearing to be at the same distance $r$ from the \gls{BS} and an average gain of one.
The \gls{BS} is assumed to be elevated enough to have no scatterers in its near field, such that the scattering is only located around the users.
Given a user $u$, the multipath components reach the \gls{BS} with normally distributed angles with mean $\varphi_{u}$ and variance $\sigma_{\varphi}^2$.
For small enough $\sigma_{\varphi}$, a valid approximation of the channel covariance matrix is $\Cm_u \in \CC^{N_r \times N_r}$ with components
\begin{equation}
\label{eq:covariance}
\left[ C_u \right]_{m,n} = e^{2\pi j d(m-n)\sin(\varphi_u)} e^{-\frac{\sigma_{\varphi}^2}{2}(2\pi d(m-n)\cos(\varphi_u))^2}
\end{equation}
%
where $d$ is the antenna spacing measured in multiples of the wavelength.
For a given user $u$, a random channel vector $\hv_u \thicksim \Cc \Nc(\mathbf{0}, \Cm_u)$ is sampled by computing
\begin{equation}
\label{eq:h_vectors}
\hv_u = \Um_u \Dm_u^{\frac{1}{2}} \Um_u^H \ev
\end{equation}
where $\ev$ is sampled from $\Cc \Nc (\mathbf{0}, \Id_{N_r})$ and $\Um_u \Dm_u \Um_u^H$ is the eigenvalue decomposition of $\Cm_u$.
The \gls{SNR} of the transmission is defined by 
\begin{equation}
\label{eq:snr}
\mathrm{SNR}=\frac{\mathbb{E}\left[\frac{1}{N_{r}}\|\mathbf{y}\|_{2}^{2}\right]}{\sigma^{2}}=\frac{1}{\sigma^{2}}
\end{equation}

\subsection{Simulation setting}
\label{subsec:simulation_setting}

\begin{figure}[!t]
\begin{tikzpicture}
	  \begin{axis}[
	    grid=both,
	    grid style={line width=.01pt, draw=gray!10},
	    major grid style={line width=.2pt,draw=gray!50},
	    xlabel={$\varphi_i$ for $i = 1,\dots,6$ (degrees)},
	    ytick={1,2,3,4,5,6,7,8,9,10},
	    ylabel={Drop number},
	    width = 239pt,
	    height = 140pt,
	  ]
	    \addplot[only marks, blue, mark=*] table [x=angle_1, y=idx_1, col sep=comma] {figs/set_angles.csv};
	    \addplot[only marks, blue, mark=*] table [x=angle_2, y=idx_2, col sep=comma] {figs/set_angles.csv};
	    \addplot[only marks, blue, mark=*] table [x=angle_3, y=idx_3, col sep=comma] {figs/set_angles.csv};
	    \addplot[only marks, blue, mark=*] table [x=angle_4, y=idx_4, col sep=comma] {figs/set_angles.csv};
	    \addplot[only marks, blue, mark=*] table [x=angle_5, y=idx_5, col sep=comma] {figs/set_angles.csv};
	    \addplot[only marks, blue, mark=*] table [x=angle_6, y=idx_6, col sep=comma] {figs/set_angles.csv};
	    \addplot[only marks, blue, mark=*] table [x=angle_7, y=idx_7, col sep=comma] {figs/set_angles.csv};
	    \addplot[only marks, blue, mark=*] table [x=angle_8, y=idx_8, col sep=comma] {figs/set_angles.csv};
	    \addplot[only marks, blue, mark=*] table [x=angle_9, y=idx_9, col sep=comma] {figs/set_angles.csv};
	    \addplot[only marks, blue, mark=*] table [x=angle_10, y=idx_10, col sep=comma] {figs/set_angles.csv};
	    
		\end{axis}
		\end{tikzpicture}
	    \caption{Ten randomly generated user drops}
	    \label{fig:users_drops}
\end{figure}
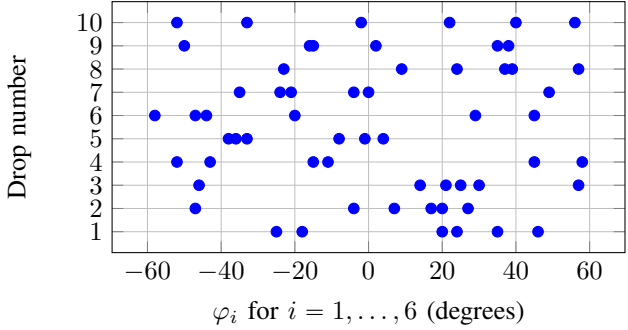

The number of antennas that equip the \gls{BS} was set to $N_r = 12$, and the number of users to $N_u = 6$.
\Gls{QPSK} modulation was considered.
The standard deviation of the multipath angle distribution $\sigma_{\varphi}$ was set to $10^{\circ}$, which results in highly correlated channel matrices.
The number of layers of MMNet in the HyperMIMO detector was set to $T = 5$.
The hypernetwork was made of 3 dense layers (see Fig.~\ref{fig:HG}).
The first layer had a number of units matching the number of inputs, the second layer 75 units, and the last layer a number of units corresponding to the number of parameters required by the detector.
The first two dense layers used \gls{ELU} activation functions, and the last dense layer linear activation functions.

Our experimentations revealed that training with randomly sampled user drops leads to sub-optimal results.
Therefore, HyperMIMO was trained with fixed channel statistics, i.e., fixed user positions.
If this might seem unpromising, our results show that HyperMIMO is still robust to user mobility (see Section~\ref{subsec:simulation_results}).
Moreover, our scheme only has $10 \times$ more parameters than MMNet as proposed in~\cite{MMNet}, which allows it to be quickly re-trained in the background when the channel statistics change significantly.
Note that this is different from MMNet that needs to be retrained for each channel matrix, which is considerably more computationally demanding.
Moreover, it is possible that further investigations on the hypernetwork architecture alleviate this issue.

Given a user drop, HyperMIMO was trained by randomly sampling channel matrices $\Hm$, \glspl{SNR} from the range [0,10]\si{dB}, and symbols from a \gls{QPSK} constellation for each user.
Training was performed using the Adam~\cite{Kingma15} optimizer with a batch size of 500 and a learning rate decaying from $10^{-3}$ to $10^{-4}$.

\subsection{Simulation results}
\label{subsec:simulation_results}

\begin{figure}[!t]
\begin{tikzpicture}
	  \begin{axis}[
	    ymode=log,
	    grid=both,
	    grid style={line width=.01pt, draw=gray!10},
	    major grid style={line width=.2pt,draw=gray!50},
	    minor tick num=1,
	    xtick={0, 2, 4, 6, 8, 10},
	    xlabel={SNR},
	    ylabel={SER},
	    legend style={at={(0.03, 0.03)},anchor=south west},
		ymax = 1e-1,	    
	    ymin = 2e-6,
	    legend cell align={left},
	  ]
	    \addplot[name path=mmse, blue, mark=diamond*] table [x=snr, y=mmse, col sep=comma] {figs/snr.csv};
	    \addplot[name path=oamp, violet, mark=triangle*] table [x=snr, y=oamp, col sep=comma] {figs/snr.csv};
	    \addplot[name path=hg, red, mark=*] table [x=snr, y=hg, col sep=comma] {figs/snr.csv};
	    \addplot[name path=mmnet, orange, mark=pentagon*] table [x=snr, y=mmnet, col sep=comma] {figs/snr.csv};
	    \addplot[name path=ml, black, mark=square*] table [x=snr, y=ml, col sep=comma] {figs/snr.csv};






	  \addlegendentry{LMMSE}
	  \addlegendentry{OAMPNet}
	  \addlegendentry{HyperMIMO}
	  \addlegendentry{MMNet}
	  \addlegendentry{Max. Likelihood}
		\end{axis}
		\end{tikzpicture}
	    \caption{\gls{SER} achieved by different schemes}
	    \label{fig:snr}
\end{figure}
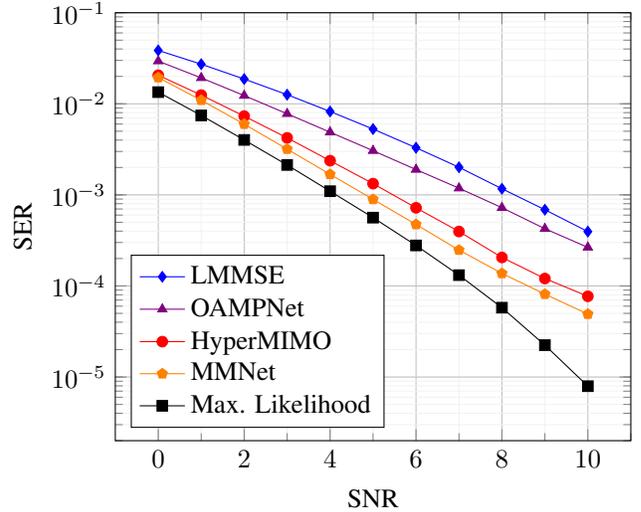

All presented results were obtained by averaging over 10 randomly generated drops of 6 users, shown in Fig.~\ref{fig:users_drops}. 
Fig.~\ref{fig:snr} shows the \gls{SER} achieved by HyperMIMO, \gls{LMMSE}, OAMPNet with 10 iteration, MMNet with 10 iterations and trained for each channel realization, and the maximum likelihood detector.
As expected, MMNet when trained for each channel realization achieves a performance close to that of maximum likelihood.
One can see that the performance of OAMPNet are close to that of LMMSE on these highly correlated channels.
HyperMIMO achieves \gls{SER} slightly worse than MMNet, but outperforms OAMPNet and \gls{LMMSE}.
More precisely, to achieve a \gls{SER} of $10^{-3}$, HyperMIMO exhibits a loss of 0.65\si{dB} compared to MMNet, but a gain of 1.85\si{dB} over OAMPNet and 2.85\si{dB} over LMMSE.

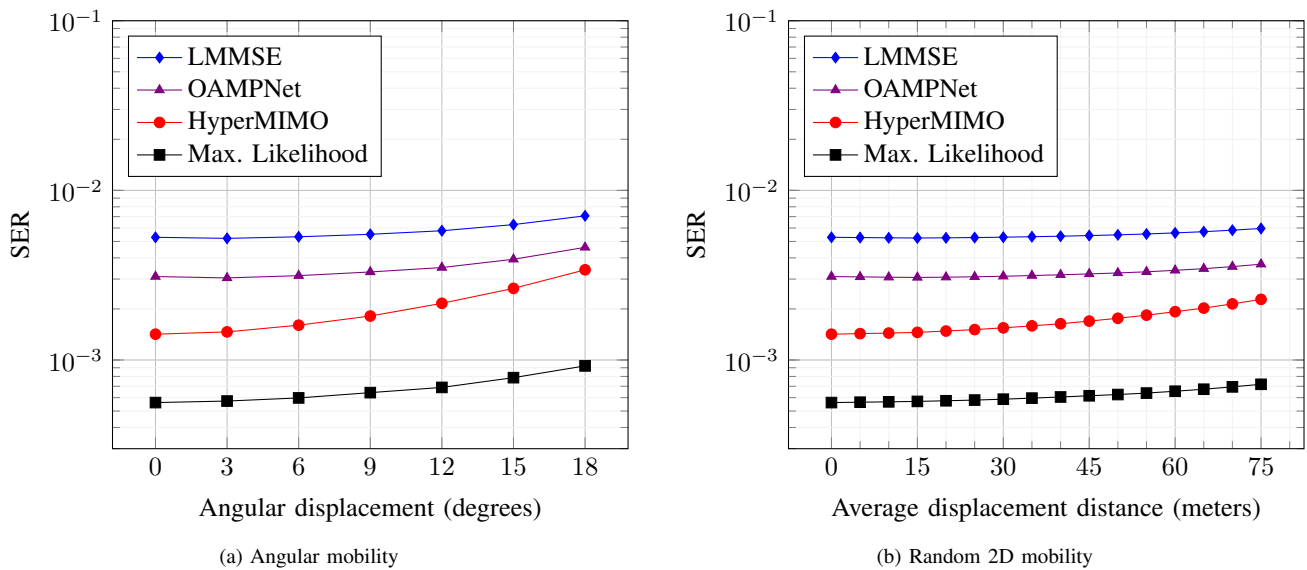
\begin{figure*}
    \centering
    \begin{subfigure}{0.45\linewidth}
		\begin{tikzpicture}
			\begin{axis}[
	    ymode=log,
	    grid=both,
	    grid style={line width=.01pt, draw=gray!10},
	    major grid style={line width=.2pt,draw=gray!50},
	    minor tick num=0,
	    xtick={0, 3, 6, 9, 12, 15, 18},
	    xlabel={Angular displacement (degrees)},
	    ylabel={SER},
	    legend style={at={(0.03, 0.635)},anchor=south west},
		ymax = 1e-1,	    
	    ymin = 3e-4,
	    legend cell align={left},
	  ]
	    \addplot[blue, mark=diamond*] table [x=angles, y=mmse, col sep=comma] {figs/angles_18.csv};
	    \addplot[violet, mark=triangle*] table [x=angles, y=oamp, col sep=comma] {figs/angles_18.csv};
	    \addplot[red, mark=*] table [x=angles, y=hg, col sep=comma] {figs/angles_18.csv};
	    \addplot[black, mark=square*] table [x=angles, y=ml, col sep=comma] {figs/angles_18.csv};

	  \addlegendentry{LMMSE}
	  \addlegendentry{OAMPNet}
	  \addlegendentry{HyperMIMO}
	  \addlegendentry{Max. Likelihood}
		\end{axis}
		\end{tikzpicture}
		\caption{Angular mobility}
		\label{fig:angles}
	 \end{subfigure} \qquad
    \begin{subfigure}{0.45\linewidth}
	    \begin{tikzpicture}
		  \begin{axis}[
	    ymode=log,
	    grid=both,
	    grid style={line width=.01pt, draw=gray!10},
	    major grid style={line width=.2pt,draw=gray!50},
	    minor tick num=2,
	    xtick={0, 15, 30, 45, 60, 75},
	    xlabel={Average displacement distance (meters)},
	    ylabel={SER},
	    legend style={at={(0.03, 0.635)},anchor=south west},
		ymax = 1e-1,	    
	    ymin = 3e-4,
	    legend cell align={left},
	  ]
	    \addplot[blue, mark=diamond*] table [x=meters, y=mmse, col sep=comma] {figs/meters_75.csv};
	    \addplot[violet, mark=triangle*] table [x=meters, y=oamp, col sep=comma] {figs/meters_75.csv};
	    \addplot[red, mark=*] table [x=meters, y=hg, col sep=comma] {figs/meters_75.csv};
	    \addplot[black, mark=square*] table [x=meters, y=ml, col sep=comma] {figs/meters_75.csv};

	  \addlegendentry{LMMSE}
	  \addlegendentry{OAMPNet}
	  \addlegendentry{HyperMIMO}
	  \addlegendentry{Max. Likelihood}
		\end{axis}
		\end{tikzpicture}
	    \caption{Random 2D mobility}
	    \label{fig:distance}
	\end{subfigure}

	\caption{\gls{SER} achieved by the compared approaches under mobility}
	\label{fig:mobility}

\end{figure*}

The robustness of HyperMIMO to user mobility was tested by evaluating the achieved \gls{SER} when users undergo angular mobility (Fig.~\ref{fig:angles}) or move in random 2D directions (Fig.~\ref{fig:distance}) from the positions for which the system was trained.
Fig.~\ref{fig:angles} was generated by moving moving all users by a given angle, and evaluating HyperMIMO for these new users positions (and therefore new channel spatial correlation matrices) without retraining.
Note that averaging was done over the two possible directions (clockwise or counterclockwise) for each user.
One can see that the \gls{SER} achieved by HyperMIMO gracefully degrades as the angular displacement increases, and never get worse thant \gls{LMMSE} nor OAMPNet.

Fig.~\ref{fig:distance} was generated by randomly moving the users in random 2D directions.
Users were located at an initial distance of $r=250$\si{m}.
The \gls{SER} was computed by averaging over 100 randomly generated displacements.
As in Fig.~\ref{fig:angles}, the \gls{SER} achieved by HyperMIMO gracefully degrades as the displacement distance increases.
These results are encouraging as they show that, despite having being trained for a particular set of user positions, HyperMIMO is robust to mobility.

\balance

\section{Conclusion}
\label{sec:conclusion}

This work proposed to leverage the recent idea of hypernetworks to alleviate the need for retraining \gls{ML}-based \gls{MIMO}-detector for each channel realization, while still achieving competitive performance.
The proposed system, referred to as HyperMIMO, uses a variation of the state-of-the-art MMNet detector~\cite{MMNet}.
To reduce the complexity of the hypernetwork, MMNet was modified to decrease its number of trainable parameters, and a form of weights sharing was leveraged.
Simulations revealed that HyperMIMO achieves near state-of-the-art performance under highly correlated channels when trained on fixed user positions.
We also show that its performance degrades slowly under user mobility, indicating that it is sufficient to re-train our scheme in the background when the channel statistics change significantly.

\balance

\bibliographystyle{IEEEtran}
\bibliography{IEEEabrv,bibliography}

\end{document}